# Manipulation of magnetizations by spin-orbit torques

*Yucai Li[1,2], Kevin William Edmonds[3], Xionghua Liu[1,2], Houzhi Zheng[1,2] and Kaiyou Wang[1, 2, 4]**

[1]*State Key Laboratory of Superlattices and Microstructures, Institute of Semiconductors, Chinese Academy of Sciences, Beijing 100083, P. R. China*

[2]*College of Materials Science and Opto-Electronic Technology, University of Chinese Academy of Science, Beijing 100049, P. R. China*

[3]*School of Physics and Astronomy, University of Nottingham, Nottingham NG7 2RD, United Kingdom.*

[4]*CAS Center for Excellence in Topological Quantum Computation, University of Chinese Academy of Science, Beijing 100049, P. R. China*

## Abstract:

The control of magnetization by electric current is a rapidly developing area motivated by a strong synergy between breakthrough basic research discoveries and industrial applications in the fields of magnetic recording, magnetic field sensors, spintronics and nonvolatile memories. In recent years, the discovery of the spin-orbit torque has opened a spectrum of opportunities to manipulate the magnetization efficiently. This article presents a review of the historical background and recent literature focusing on spin-orbit torques (SOTs), highlighting the most exciting new scientific results and suggesting promising future research directions. It starts with an introduction and overview of the underlying physics of spin-orbit coupling effects in bulk and at interfaces, then describes the use of SOTs to control ferromagnets and antiferromagnets. Finally, we summarize the prospects for the future developments of spintronics devices based on SOTs.

## 1 Introduction

In 1996, Slonczewski [1] and Berger [2] independently predicted that electrical current can reverse the magnetization of magnets through the direct action of the current spin polarization on the local magnetic moments, an effect known as spin transfer torque (STT). Since then, magnetization control by electric current has attracted increasing attention. For STT, the charge current flowing across the ferromagnetic layer induces the spin polarization, which affects ferromagnetic magnetization through the interactions between the spin polarized electrons and local magnetic moments [3-4]. STT has been used to develop magnetic random access memory (MRAM) based on magnetic tunneling junctions (MTJ) [5], where it offers advantages over competing technologies due to its high speed and non-volatility [6]. However, the requirement to write information by passing a high current density directly through an insulating tunnel barrier impacts upon its stability.

As an alternative to STT, manipulation of magnetization by spin-orbit torques (SOTs) could enable faster and more stable devices with lower power consumption [7-10]. The SOTs originate from the strong spin-orbit coupling [11], which can transfer charge current into spin current in non-magnetic semiconductors or heavy metals [12-13]. In 2009, current-assisted magnetization switching by SOTs was first observed in the ferromagnetic semiconductor GaMnAs [14]. Since then, SOT effects on magnetization by SOTs have been extended to ferromagnetic metal multilayers and antiferromagnets [15]. Taking advantage of a lateral wedge oxide [16], a ferroelectric substrate [17], and interlayer exchange coupling by an ferromagnetic layer [18] or an antiferromagnetic layer [19-20], current-induced magnetization switching has been demonstrated without external magnetic field. Furthermore, SOTs can also be utilized to drive non-uniform magnetization states including domain walls [21] and skyrmions [22].

In this review, firstly the underlying physics and the different kinds of spin-orbit couplings in bulk and interface are briefly introduced. We then review the recent progress of the SOTs on ferromagnets and antiferromagnets. Finally, we point to the prospects for future developments of spintronics devices based on SOTs.

## 2 Spin-orbit coupling in materials

Spin-orbit coupling (SOC) is a relativistic effect of a particle's spin with its motion in an electric field. In atomic energy levels, the SOC can split degenerate states with finite angular momentum (*p*, *d*, and *f*), and the contribution to the Hamiltonian can be expressed as [23]

$$E_{SO} = \frac{Z^4}{2(137)^2 a_0^3 n^3}\left(\frac{j(j+1)-l(l+1)-s(s+1)}{2l(l+1/2)(l+1)}\right),$$

where $l$, $s$ and $j$ are the orbital, spin and total angular momentum quantum numbers, respectively, $Z$ is the effective nuclear charge and $n$ is the principal quantum number. It is noticed that the SOC increases as the fourth power of $Z$ but as the third power of $n$, suggesting the SOC is larger for atoms that are further down a particular column of the periodic table.

Although the SOC term contributes a small perturbation to the Hamiltonian, it can influence the energy band structure and Fermi surface effectively in solids. Inversion symmetry breaking, whether due to underlying crystal structure or interfaces, plus SOC causes new terms in the Hamiltonian of the charge carriers which are antisymmetric in carrier momentum. Depending on the nature of the symmetry breaking, different types of SOC effects have been observed in the solid state, including Dresselhaus effect, Rashba effect, Rashba-Edelstein effect, and Dzyaloshinskii-Moriya interaction (DMI).

The Dresselhaus SOC originates from the bulk inversion asymmetry, as was first discovered by Dresselhaus [24] in non-centrosymmetric crystalline zinc-blende III-V semiconductors. The form of cubic Dresselhaus Hamiltonian can be written down as [24-25]:

$$H_D = (\gamma / \hbar)((p_y^2 - p_z^2) p_x \sigma_x + c.p.)$$

where $\gamma$ is the gyromagnetic ratio, $\hbar$ is the reduced Planck constant, $p_x$, $p_y$, $p_z$ are the components of the momentum in the [100], [010] and [001] directions, respectively, $\sigma$ are the Pauli matrices for the spin and $c.p.$ denotes circular permutations of indices.

Further breaking of symmetry, such as the deformation introduced by uniaxial strain, will generate additional SOC terms [26]. Assuming the strain is along [001], the cubic Dresselhaus SOC gives way to linear Dresselhaus SOC [25]:

$$H_D = (\gamma p_z^2 / \hbar)(p_x \sigma_x - p_y \sigma_y)$$

At the interfaces or surfaces of materials, the lack of spatial inversion symmetry generates a built-in internal electric field, and the itinerant electrons experience an effective magnetic field, called the Rashba SOC [27], with the corresponding Hamiltonian [25, 28]:

$$H_R = (\alpha_R / \hbar)(p_x \sigma_y - p_y \sigma_x)$$

where $\alpha_R$ is the Rashba parameter.

Both Dresselhaus and Rashba effects can induce spin-momentum locking and energy band splitting, and the resulting spin textures at the Fermi surface are presented in **figure 1a and 1b**, respectively. The spin-momentum locking from surface or interface can transfer charge current to spin current, known as Rashba-Edelstein effect or Edelstein effect [25, 29], which is shown in **figure 1c and 1d**.

Furthermore, for materials lacking inversion symmetry, the SOC competing with the exchange interaction results in a chiral exchange interaction referred to as the DMI [28]. This effect was first introduced to describe the weak ferromagnetism of antiferromagnetic insulators [30-31] and subsequently extended to non-centrosymmetric magnetic metals [32] and magnetic multilayers [33-34]. The Hamiltonian of the DMI is $\boldsymbol{D}_{ij} \cdot (\boldsymbol{S}_i \times \boldsymbol{S}_j)$ [30-31, 35], where $\boldsymbol{S}_i$ and $\boldsymbol{S}_j$ denote the neighbour spins, and the vector $\boldsymbol{D}_{ij}$ $(= -\boldsymbol{D}_{ji})$ depends on details of electron wave functions and the symmetry of the crystal structure [32, 36]. The illustration of DMI in the bilayer is displayed in the **figure 1e.**

The spin Hall effect (SHE) and anomalous Hall effect (AHE) are closely related transport phenomena resulting from SOC. When electric current passes through a heavy metal or semiconductor with strong SOC, opposite sign spins will be accumulated at opposing lateral surface boundaries [37]. This effect, the SHE, was first predicted by Dyakonov and Perel in 1971 [38]. More than thirty years later, Awschalom *et al.* [39] and Wunderlich *et al.* [40] independently observed the SHE experimentally.

The SHE can be of either extrinsic and/or intrinsic origin. The extrinsic mechanism is due to electrons scattering against spin-orbit coupled impurities (Mott scattering) [38]. The other mechanism is from intrinsic band structure properties of the materials: the SOC can distort the electron trajectories and results in the SHE [40], which is illustrated in **figure 1f.** The strength of SHE can be quantified using the spin Hall angle, which is defined by the ratio of the spin Hall conductivity against charge conductivity.

In ferromagnetic materials, in addition to the ordinary Hall resistance, a large anomalous Hall resistance [41-42] is often observed, which is dependent on the magnetization and caused by the SOC. Similar to the SHE, the origin of the AHE has extrinsic and intrinsic mechanisms: the former can have contributions from skew scattering [43] and side-jump scattering [44] due to the impurities, while the latter is from the band structure effect [41]. AHE provides an effective way to detect the magnetization of materials electrically.

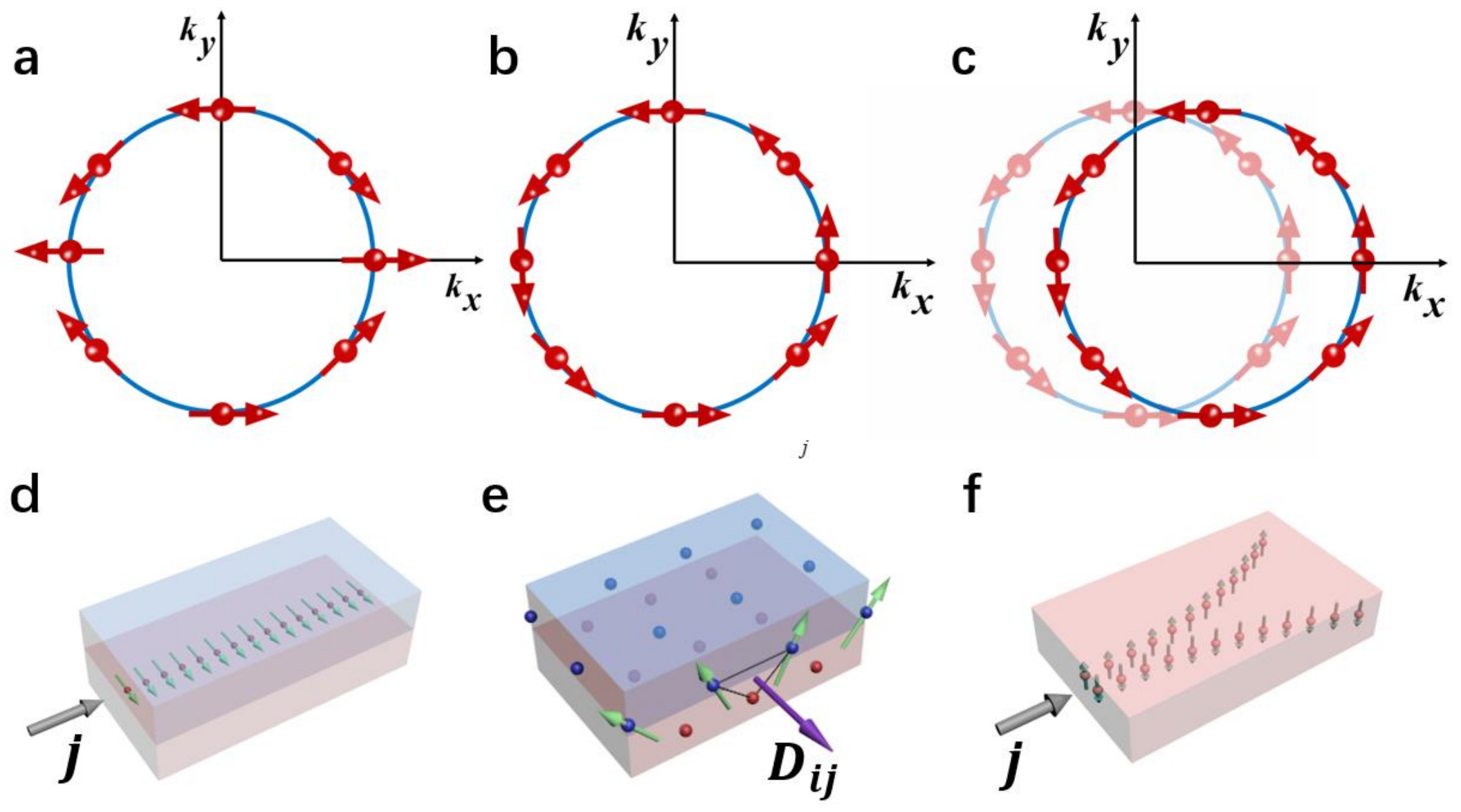

**Figure1. Illustration of spin-orbit coupling phenomena. a,b,** Spin texture induced by Dresselhaus SOC when strain is along [001] and Rashba SOC in the momentum space. **c**, The nonequilibrium spin distribution due to electrical current and the Rashba-Edelstein effect. **d**, Schematic of the Rashba-Edelstein effect. **e,** Schematic of DMI at interface of bilayer. **f,** Schematic of spin Hall effect.

## 3 Manipulation of magnetic materials by spin orbit torques

### 3.1 Current induced ferromagnets switching by SOTs

#### 3.1.1 SOTs in the ferromagnets and magnetization dynamics

Current-assisted switching by SOTs was first discovered in $Ga_{1-x}Mn_xAs$ by Chernyshov *el al.* [14] in 2009. The spin-orbit field generated by electrical current was shown to enable reversible manipulation of the magnetization between two orthogonal easy axes.

Later, detailed studies of the anisotropy of current-induced switching [45-46] as well as spin-orbit driven ferromagnetic resonance [47] have provided a quantitative determination of the relative strength of Dresselhaus-type and Rashba-type SOTs in GaMnAs.

Soon after Chernyshov *el al.*'s work, investigations of SOTs were extended to magnetic heterostructures. Miron *et al.* reported the domain wall motion driven by SOTs [48] and the magnetization switching by electric current [49-50]. The original interpretation was suggested that the Rashba effect can yield a large effective magnetic field in a $Pt/Co/AlO_x$ ferromagnetic heterostructure, where the $AlO_x$ top layer and the Pt bottom layer maximize the structural asymmetry of the system [48-50]. The orbital angular momentum from the crystal lattice is transformed to the local spin magnetization induced by the Rashba SOC, thereby provoking the reversal of magnetization [49-50].

However, Liu *et al.* [51] ascribed the magnetization switching in a CoFeB/β-Ta bilayer to the SHE rather than the Rashba effect, where the spin-polarized current in the adjacent heavy-metal layer transfers spin angular momentum to the ferromagnetic CoFeB, resulting in a spin torque. Moreover, Liu *et al.* [52] also reported the current-induced magnetization switching in $Pt/Co/AlO_x$ by the SHE, where they concluded that the Rashba effect offers no measurable contribution to the switching.

Generally, to make clear the contributions from Rashba effect and/or SHE to SOTs is a crucial issue, which can provide a useful physical guidance to the application. However, up to now, the relative importance of the SOTs induced by the SHE or Rashba effect remains ambiguous though many efforts have been devoted to understanding the underlying mechanism [42, 53-55].

SOTs switching of ferromagnetic layers with in-plane [51, 56] and out-of-plane [50-51, 57] easy axes have both been widely studied, with a qualitatively similar switching mechanism in each case. In this review, we mainly focus on the SOTs in the materials with perpendicular magnetic anisotropy, which are the subject of most researches and applications.

The action of SOTs on magnetic materials can be described by adding an extra term in the Landau-Lifshitz-Gilbert (LLG) equation describing precessional magnetization dynamics:

$$\frac{d\boldsymbol{M}}{dt} = -\gamma \boldsymbol{M} \times \mathbf{H}_{eff} + \frac{\alpha}{M_s} \cdot (\boldsymbol{M} \times \frac{d\boldsymbol{M}}{dt}) + \frac{\gamma}{\mu_0 M_s} \boldsymbol{\tau}_{SOT}$$

where the $\gamma$ is the gyromagnetic ratio, $\mathbf{H}_{eff}$ is the effective magnetic field (sum of the external field as well as anisotropy and exchange fields), $\alpha$ is the Gilbert damping coefficient, $\boldsymbol{\tau}_{SOT}$ is the SOT. The SOT term can consist of both a field-like torque $\boldsymbol{\tau}_{FL} \sim \boldsymbol{m} \times \boldsymbol{y}$ and a damping-like torque $\boldsymbol{\tau}_{DL} \sim \boldsymbol{m} \times (\boldsymbol{y} \times \boldsymbol{m})$, where $\boldsymbol{m}$ is the magnetization unit vector and $\boldsymbol{y}$ denotes the axis perpendicular to the current in plane. Both field-like torque and damping-like torque can be generated by Rashba and SHE but the SOTs from different mechanisms differ in magnitude [55]. It is well-known that magnetization switched by STT undergoes many precessions due to the competition between damping torque and damping-like terms of STT. Unlike the STT, the damping-like term in SOT is perpendicular to the damping torque thus does not compete with damping torque directly, resulting in faster magnetization switching [58-60].

The torques for current-induced magnetization switching in heavy-metal (HM)/ferromagnet (FM) bilayer are illustrated in **figure 2**.

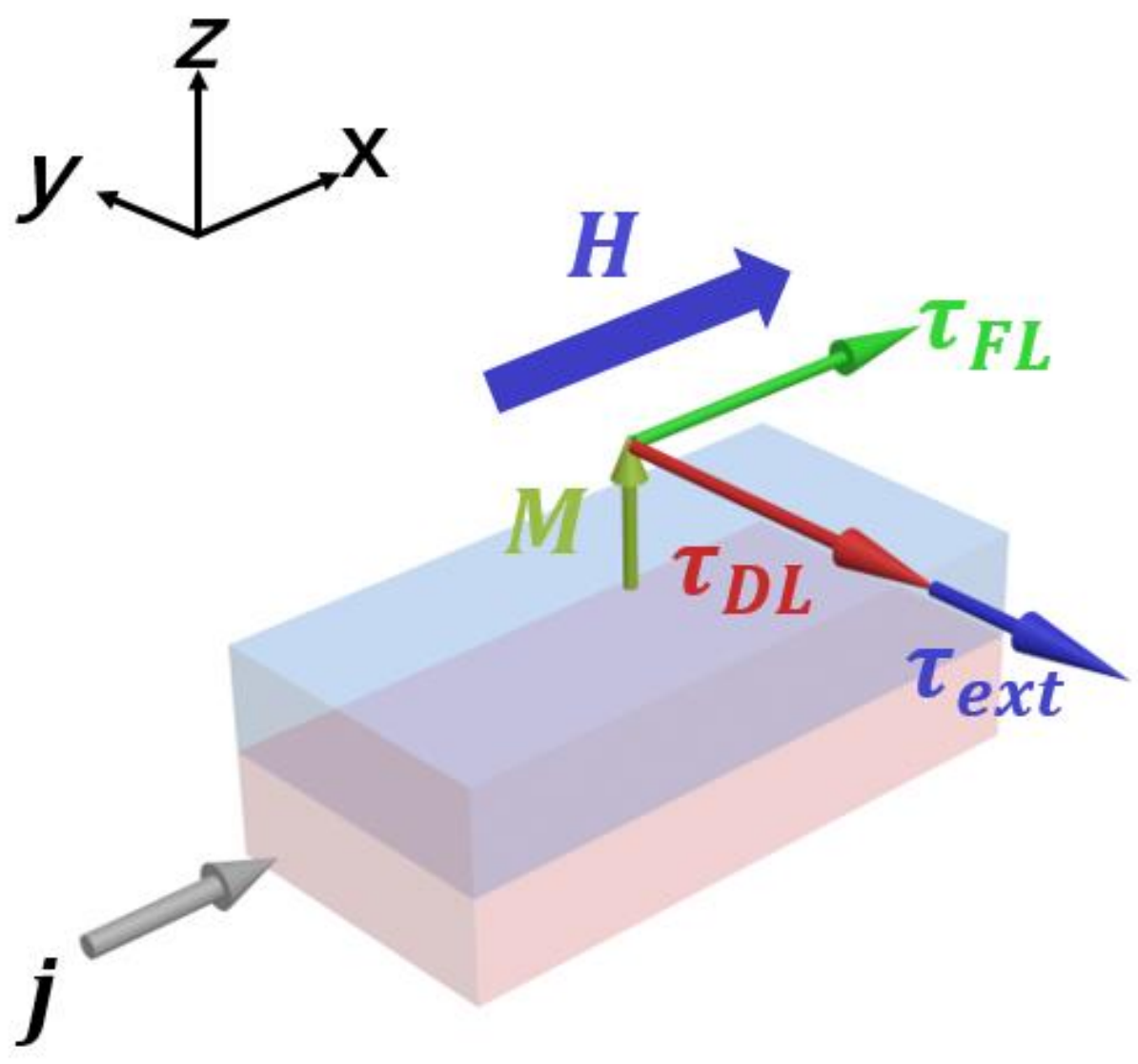

**Figure 2.** Schematics of the magnetic torques in a HM/FM bilayer with external field **H** and current density ***j***. $\boldsymbol{\tau}_{FL}$ is the field-like SOT and $\boldsymbol{\tau}_{DL}$ is the damping-like SOT. $\boldsymbol{\tau}_{ext}$ is the torque induced by the external magnetic field.

SOTs can be used in switching the magnetic free layer of a MTJ, which forms the memory cell of a MRAM [61]. The parallel or antiparallel alignment of magnetization between the free layer and the reference layer in MTJ corresponds to the different states of tunneling magnetoresistance (TMR), as illustrated in **figure 3**. Three-terminal SOT-MTJ, with separated channels for writing and reading operations using SOTs and TMR respectively, have many advantages for application: low power consumption [62], ultra-fast writing (sub-ns reversal) [63] and high-reliability [64]. In addition, combining the SOTs and STT can overcome major drawbacks of the conventional STT-MRAM and SOT-MRAM [65]. Erasing by SOTs and programming by STT offer a new path to design spin memory devices.

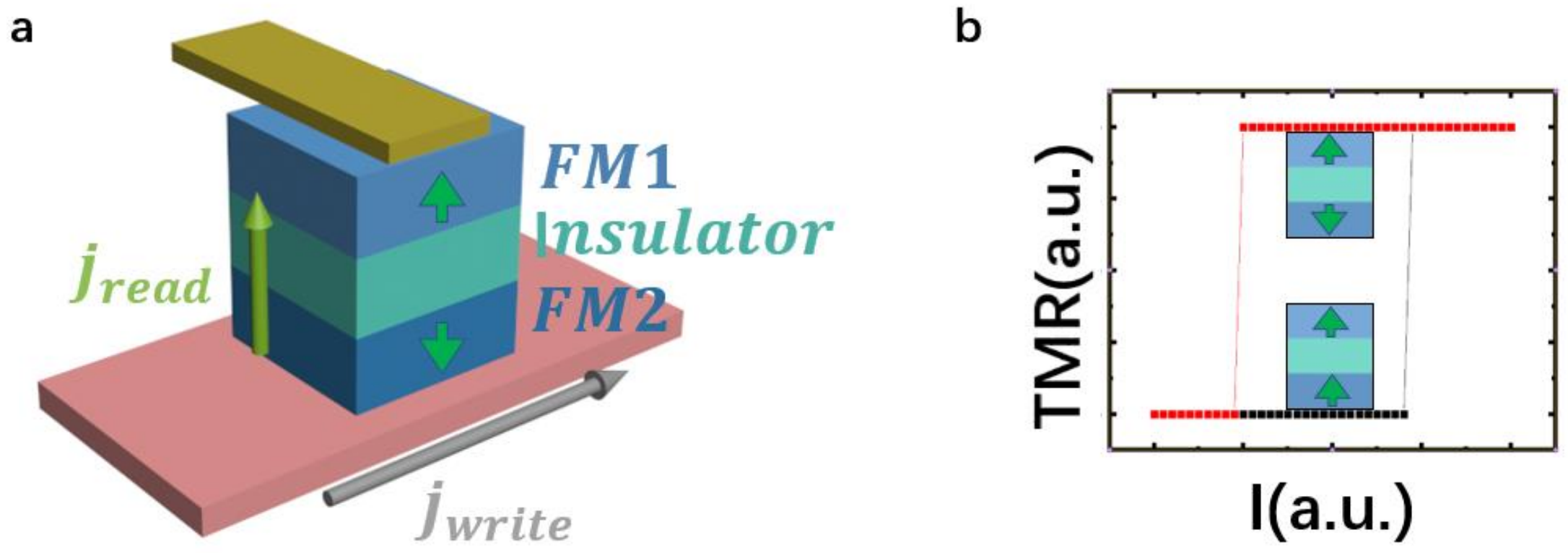

**Figure3 Schematics of SOT-MTJ. a,** The writing process and reading process of MTJ. Writing current is applied in bottom HM layer and switches the FM2 layer. A low current is applied across the junction to read the states of MTJ. **b**, Schematic of the dependence of TMR on the states of SOT-MTJ. The current can switch the magnetization, as a result, the TMR changes with the current.

In addition, SOTs can be used to control spin waves. The ferromagnetic insulator $Y_3Fe_5O_{12}$ (YIG) is a favoured material for such studies as its spin-wave damping factor is the lowest in nature. In 2010, Kajiwara et al. [66] reported the generation and detection of spin waves by SOTs in YIG/Pt structures. A charge current in the Pt layer induces, via the SHE, a spin angular momentum which can propagate through the YIG before being detected at another Pt film ~1mm away. Hamadeh *et al.* [67] demonstrated that the magnetic losses of spin-wave modes can be reduced or enhanced by at least a factor of 5 depending on the magnitude and polarity of the electrical current in Pt. The damping of the fundamental mode could be completely compensated by the SOT, and coherent auto-oscillation was achieved [68]. Tuning of the spin-wave damping and SOT-induced auto-oscillation has also been achieved in ferromagnetic metallic multilayers Ta/CoFeB [69] and Pt/NiFe[69] [70].

### 3.1.2 Current-induced domain-wall or skyrmions motion by SOTs

A magnetic field applied to a multi-domain ferromagnet will favor the expansion of a magnetic domain over its neighbors and move the bordering domain walls [71-73]. The field-induced domain wall motion usually has no preferred direction. However, a spin-polarized current can move a domain wall in the direction of the electron, independent of the polarity of the wall, so that current-driven domain wall motion has been proposed for memory and logic devices [74-76].

In 2010, Miron *et al.* [48] reported the domain-wall motion driven by SOTs. The domain wall velocity can reach up to 400 m $s^{-1}$ in an ultrathin Pt/Co/$AlO_x$ nanowire. The asymmetric layer structure can induce large DMI [77]. Thiaville *et al.* [78] proposed that for Néel domain walls stabilized by the DMI in ultrathin magnetic films, current-induced wall motion by SOTs can be efficient because the magnetization direction of Néel domain walls is orthogonal to the direction of spin polarization by the SHE (see **figure 4**) . Moreover, Emori *et al.* [79-80] demonstrated the dynamics of chiral ferromagnetic domain walls driven by the SOTs in Pt/CoFe/MgO and Ta/CoFe/MgO structures. The opposite movements of domain walls with electric current flowing in Pt/CoFe/MgO and in Ta/CoFe/MgO suggest the effective SOTs driving the domain walls with opposite directions. Similar work was also independently reported by Ryu *el al.* [81].

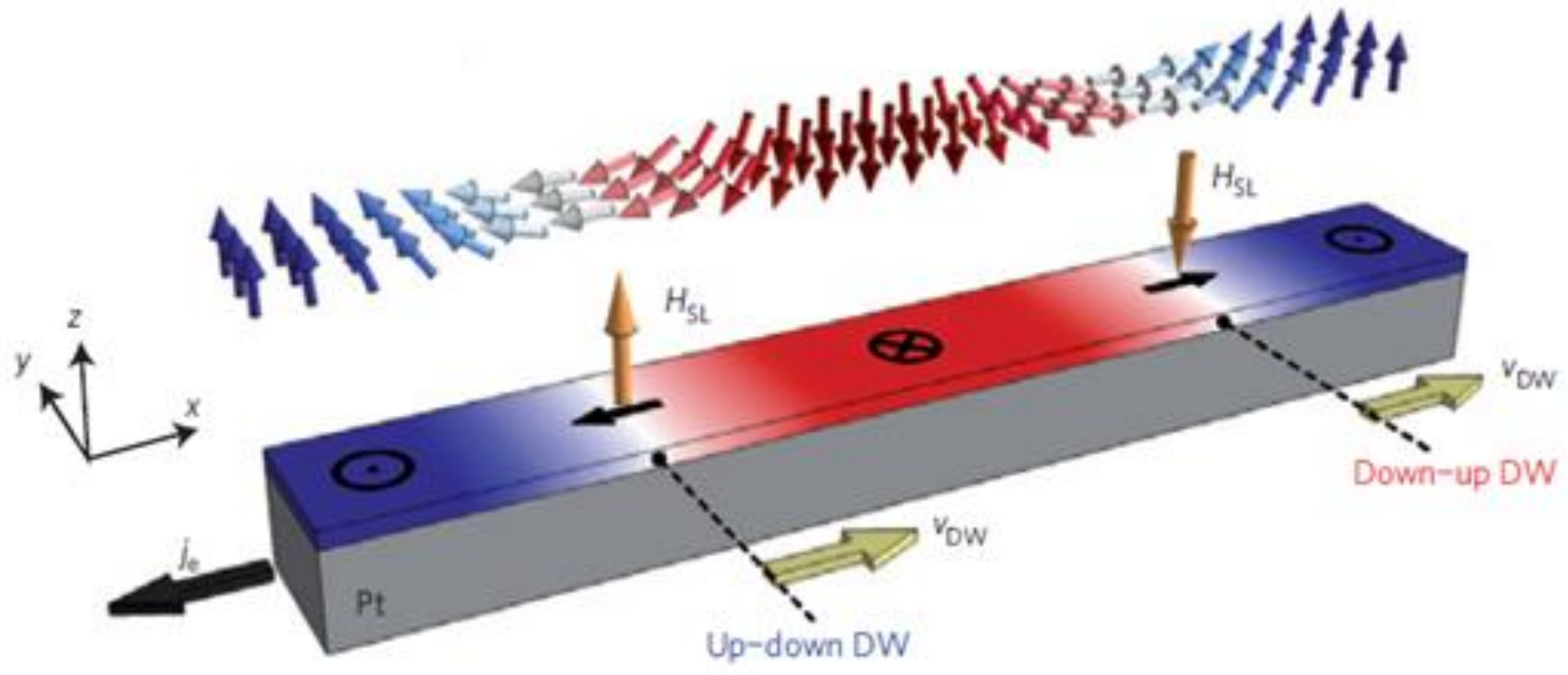

**Figure 4**. Dynamics of current-driven domain wall motion. The Néel domain walls in Pt/CoFe/MgO have left-hand chirality and the current drives the domain walls' motion in the same direction. Reproduced with permission.[79] Copyright 2013, Nature Publishing Group.

Generally, SOTs in HM/FM structure drive the domain walls in the same direction with same speed, without changing the overall magnetization. However, when an external magnetic field applied along the current direction alters the central domain wall moments, the SOTs driving the adjacent domain walls with different velocities will result in the magnetization switching [21, 79, 82-83]. Recently, magnetization switching by SOT-driven domain wall motion has been observed by spatially and time-resolved magneto-optical Kerr effect (MOKE) [82].

Driving the motion of skyrmions by SOTs is attractive due to the ultra-low critical current density (below $10^2$ Acm$^{-2}$) [84] and potential for high density information storage. In 2012, the motion of skyrmions driven by STT was observed in an FeGe microdevice [84]. Subsequently, the local nucleation of skyrmions by spin-polarized current generated by the SHE was reported [85]. As the skyrmions move along a track due to the SOTs, their velocities have longitudinal and transverse components. The longitudinal motion depends on the chirality of the skyrmions and the spin Hall angle, while the transverse one depends on the direction of magnetization in the center of the skyrmions. The transverse motion is often called skyrmion Hall effect. When the skyrmions are close to the edge, the motion is parallel to the track due to the repulsive force exerted by the edge [85-86].

Current-induced skyrmion motion was observed by MOKE microscopy in a Ta/CoFeB/$TaO_x$ heterostructure [87]. Above a critical current density due to the pinning by defects, both the longitudinal and transverse components of velocity were clearly evident [87]. Similar phenomenon was also observed in Pt/CoFeB/MgO [88], where the velocity linearly varies with the current, in agreement with magnetic simulations. However, some features of the motion are not consistent with theoretical expectations, such as the dependence of the velocity and the size of skyrmion, [22, 86-87] and the relation between skyrmion Hall angle and the velocity [86, 89]. Therefore, the underlying mechanism remains controversial [86].

The skyrmion Hall effect hinders the practical application of skyrmion in racetrack memory. The opposite transverse components of the coupled skyrmions motion in different magnetic sublattices can result in their cancellation, leading to pure longitudinal motion, which was studied theoretically in antiferromagnetic systems [90-92].

### 3.1.3 Field-free deterministic magnetization switching by SOTs.

In perpendicularly magnetized HM/FM structures, magnetization can be switched by SOTs to a fixed direction under an in-plane magnetic field; without a magnetic field, switching is to a random up or down direction. The requirement for an external magnetic field greatly limits the applications of SOT. Therefore, a large volume of work has focused on magnetization switching by SOTs without external field. Importantly, in 2014 Yu *et al.* [16] achieved field-free deterministic magnetization switching in a $Ta/Co_{20}Fe_{60}B_{20}/TaO_x$ structure by introducing a lateral structural asymmetry. In their work, a wedge-shaped $TaO_x$ layer produces a field-like SOT due to the Rashba effect, with an effective magnetic field along the z axis that facilitates current-induced deterministic switching. Thus the asymmetrical geometric shape provides an effective means to achieve deterministic switching [93-94]. However, the requirement for a lateral asymmetry presents problems for practical applications due to the difficulties in fabrication. In 2016, field-free magnetization switching was realized by utilizing the exchange bias effect due to an adjacent antiferromagnetic layer [20]. The interlayer exchange coupling due to Ruderman–Kittel–Kasuya–Yosida interaction between two FM layers has been also used to enable field-free magnetization switching [18]. In this study, the symmetry of two states of perpendicular magnetization was broken due to the exchange coupling to a layer with in-plane magnetization. Thus the deterministic switching could be achieved, and moreover could be tuned by switching the in-plane magnetization direction [18].

Cai *et al.* [17] reported that the polarization of PMN-PT ferroelectric substrates can be used to control the spin-orbit torque in the absence of an external magnetic field in a PMN-PT/Pt/CoNiCo/Pt system, as shown in **figure 5a**. Before the substrate is polarized, the magnetization deterministically switches to a fixed z direction regardless of whether the current is applied along the x or -x direction. After applying a voltage of +500 V to polarize the ferroelectric substrate along the x axis, a clockwise hysteretic loop of magnetization versus current density curve was reproducibly observed **(figure 5b**). The loop then becomes anticlockwise after applying a voltage of -500V (see **figure 5c**). The gradient of spin densities caused by the polarization of the PMN-PT produces an extra torque which enables the voltage-controlled deterministic switching.

Recently, Ma *et al.* [95] presented a novel switching behavior in a Pt/W/CoFeB/MgO structure. Surprisingly, the competing spin currents in the Pt and W layers with opposite spin Hall angle produce deterministic switching without magnetic field. This interesting phenomenon is difficult to interpret with present models.

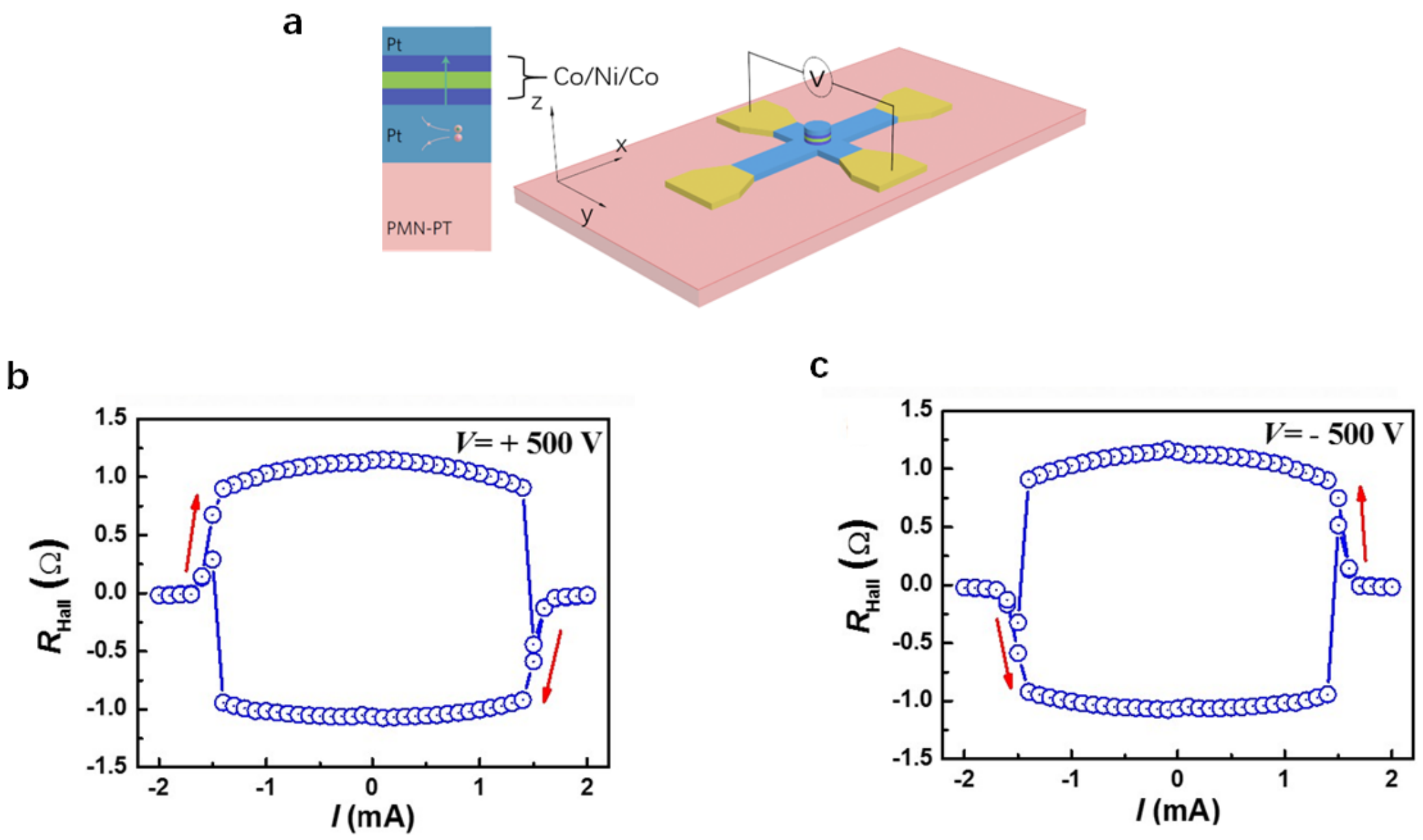

**Figure 5**. Magnetization switching induced by the polarization of PMN-PT substrate. **a,** The structure of the device; **b, c** The deterministic magnetization switching after applied 500V and -500V voltage [17]. Copyright 2017, Nature Publishing Group.

### 3.1.4 Interfacial engineering.

Modifying the interface asymmetry using, *e.g.*, spacer layers, capping layers, oxidation *etc.* provides an important route to tune the magnitude and direction of SOTs [96-98].

Fan *et al*. [96] investigated the effect of copper spacer layers between Pt and CoFeB, finding that the ratio between the damping-like torque and field-like torque has a sudden change around the 0.7 nm Cu. Moreover, the insertion of Hf layer in between Pt and CoFeB was reported to decrease the spin mixing conductance and reduce the damping [99]. Both of these studies demonstrate the importance of the interface effect on SOTs.

By controlling the thickness of the top $SiO_2$ layer in a Pt/CoFeB/MgO/$SiO_2$ heterostructure, Qiu *el al.* [100] tuned the oxidation of the CoFeB layer and thus the interface and effective field were modified. A different switching sequence was observed for small and large thicknesses of $SiO_2$, with a sign change of the effective field occurring for the thickness of 1.5 nm $SiO_2$. This means the interfacial SOTs can be effectively tuned by oxidation. Emori *et al.* [101] modified the SOTs in a Pt/Co/$GdO_x$ heterostructure by voltage-driven $O^{2-}$ ion migration. The oxidation of Co can be controlled by the voltage across the Co/$GdO_x$ interface and the damping-like torque is significantly enhanced. The oxidation state at the ferromagnet/oxide interface modified by voltage also can induce a change of the magnetization anisotropy (voltage controlled magnetic anisotropy, VCMA) [102]. Combined with VCMA, the critical current of magnetization switching by SOTs can be effectively tuned [103]. The oxidation of the HM layer also can vary the SOTs in $WO_x$/CoFeB/TaN system [16].

The capping layer is also found to greatly affect the SOTs. Qiu *et al.*

[104] observed the enhancement of SOTs using a Ru capping layer, by comparing HM/FM/Ru and HM/FM/MgO heterostructures. For the HM/FM/MgO structure, if the FM layer thickness is thinner than the spin dephasing length, the spin current injected by the HM cannot be fully absorbed by the FM layer and is reflected back at the FM/ MgO interface. For the HM/FM/Ru system, the Ru acquires a negative spin polarization which enhances the absorption of spin current.

Seung-heon *et al.* [105] reported a new mechanism for spin current generation and demonstrated field-free switching in a CoFeB/Ti/CoFeB/MgO structure, where the magnetic easy axes for the bottom and top magnetic CoFeB are in-plane and out-of-plane, respectively. The spin-orbit filtering and the spin precession induced by interfacial spin-orbit field are two distinct mechanisms for interfacial spin scattering. The former gives a transverse spin polarization and the latter generates a spin current with a z (the axis normal to the interface) component. When the current is flowing in the device along the x direction, the spin polarization of the spin current has additional z component, which facilitates the magnetization switching.

### 3.1.5 Charge to spin conversion

To increase the efficiency of magnetization switching by SOTs, it is crucial to search for materials with large spin Hall angle (the ratio between spin current density and charge current density). Metals with strong spin orbit coupling [106] such as Pt [107], Ta [51], Pd [108], W [109], Hf [110], etc have been widely studied in the past several years. It has been reported that the spin Hall angle is significantly related to the crystalline structure and can be varied by changing the thickness [111] or by oxidation [112].

The SHE is not limited to nonmagnetic materials. Miao *et al.* [113] demonstrated the inverse SHE in a ferromagnetic permalloy film, and Du *et al.* [114] determined the spin-Hall angle in a range of 3*d* transition metal films including antiferromagnetic Cr and FeMn. Zhang *et al.* [115-116] demonstrated that the 5*d* metal alloys, such as IrMn and PtMn, exhibit a larger spin Hall angle than 4*d* metal alloys, such as PdMn, and 3*d* ones, such as FeMn. Thus, in antiferromagnetic alloys, the heavy nonmagnetic elements dominate the SHE.

Topological insulators with large spin Hall angle, such as $Bi_2Se_3$ [117], have attracted increasing attention for current-induced magnetization switching. Topological insulators are insulating in the interior but have metallic states on the surface [118], due to the strong SOC causing band structure inversion [119]. The surface states of three-dimensional topological insulators have Rashba spin texture. When applying an electrical field, the current mostly flows in the surface and the spin momentum locking can produce non-equilibrium spin accumulation [120-121]. In 2014, Mellnik *et al.* [122] reported SOTs in the topological insulator/ferromagnet ($Bi_2Se_3/Ni_{81}Fe_{19}$) heterostructure due to current flowing through the topological surface state. Soon after, Fan *et al.* [123] demonstrated magnetization switching in a $(Bi_{0.5}Sb_{0.5})_2Te_3/(Cr_{0.08}Bi_{0.54}Sb_{0.38})_2Te_3$ bilayer at 1.9 K, where the Cr-doped topological insulator $(Cr_{0.08}Bi_{0.54}Sb_{0.38})_2Te_3$ is ferromagnetic. The effective field to current ratio and the spin Hall angle were detected by second-harmonic methods, and were found to be much larger than for HM/FM systems. Similar phenomena in topological insulator/ferromagnet systems were also observed by other groups [124-125]. The ultralow power SOT switching by topological insulators has been shown by NHD Khang *et al.* [126]. The $Bi_{0.9}Sb_{0.1}$ has large spin Hall angle (5200%) and conductivity ($2.5\times10^5\Omega^{-1}m^{-1}$) and the critical switching current has been lowered to $1.5\times10^6 Acm^{-2}$. The challenges for application are using industry-friendly sputtering method, rather than the molecular beam epitaxy method to fabricate BiSb. Recently, high SOTs due to quantum confinement in sputtered $Bi_xSe_{(1-x)}$ films has been demonstrated by D. Mahendra *et al.* [127], the critical switching current has been reduced to $4.3\times10^5 Acm^{-2}$. These researches indicate that topological insulators might be a good candidate to optimize the SOTs efficiency.

Oxide interfaces with strong Rashba effect have also been actively studied [128-130]. $LaAlO_3$ (LAO) and $SrTrO_3$ (STO) are both insulators, however, a conductive two-dimensional electron gas exists at the interface between LAO and STO [131]. The inverse Edelstein effect due to the symmetry breaking measured by spin pumping have shown that spin-to-charge conversion through Rashba coupling can occur in the interface[130]. Recently, Wang *et al.* [132] reported the direct charge-to-spin conversion at the STO/LAO interface and a large spin Hall angle up to 630 % was detected at room temperature, which is far higher than that of the reported heavy metals.

We summarize the key parameters describing the spin conversion efficiency obtained experimentally for different materials in Table 1, and the critical switching current density in Table 2, from which we may find a suitable material (or structure therein) to lower the critical current that is used to effectively manipulate magnetization by SOTs.

Table 1

$T$ is the temperature for the measurement with RT for room temperature, $\lambda_{Sd}$ is the spin diffusion length, $\sigma_{NM}$ is the conductivity, $\theta_{SH}$ is the spin Hall angle, Method is the measuring method, NLD = nonlocal detection, SP=spin pumping, ST-FMR= spin torque ferromagnetic resonance, SMR = Spin Hall magnetoresistance SHM= second-harmonic method.

| Materials | $T$ (k) | $\lambda_{Sd}$ (nm) | $\sigma_{NM}(10^6\Omega^{-1}m^{-1})$ | $\theta_{SH}(\%)$ | Method | Reference |
|---|---|---|---|---|---|---|
| Al (12nm) | 4.2 | 455±15 | 10.5 | 0.032±0.006 | NLD | [107, 133] |
| Al (25nm) | 4.2 | 705±30 | 17 | 0.016±0.004 | NLD | [107, 133] |
| Cr | RT | 13.3 | / | -5.1±0.5 | SP | [114] |
| Mn | RT | 10.7 | / | -0.19±0.01 | SP | [114] |
| Ti | RT | 13.3 | / | -0.036±0.004 | SP | [114] |
| V | RT | 14.9 | / | 1±0.1 | SP | [114] |
| Au | 4.5 | 65 | 48.3 | <2.3 | NLD | [134] |
| | ≤10 | 40±16 | 25 | 1.4±0.4 | NLD | [135] |
| | RT | 86±10 | 37 | 11.3 | NLD | [136] |
| | RT | 83 | 37 | 3 | NLD | [137] |
| | RT | 35±4 | 28 | 7±2 | NLD | [138] |
| | RT | 35±3 | 25.2±0.13 | 0.35±0.03 | SP | [139] |
| | RT | 35±3 | 5.26 | 1.6±0.1 | SP | [140] |
| | RT | 35±3 | 7 | 0.335±0.006 | SP | [140] |
| | RT | 35 | / | 1.1±0.3 | SP | [141] |
| $Au_{99.58}Fe_{0.42}$ | RT | 33±3 | 23.3 | 7±1 | NLD | [138] |
| $Au_{99.05}Fe_{0.95}$ | RT | 27±3 | 14.3 | 7±3 | NLD | [138] |
| $Au_{98.6}PT_{1.4}$ | RT | 25±3 | 14.5 | 12±4 | NLD | [142] |
| $Au_{98.6}Pt_{1.4}$ | RT | 50±8 | 16.7 | 0.8±0.2 | NLD | [142] |
| $Au_{93}W_{7}$ | RT | 1.9 | 1.75 | >10 | NL and SP | [143] |
| Bi (interface) | RT | / | 2.4±0.3 | -(7.1±0.8) | SP | [144] |
| Bi(volume) | RT | / | 50±12 | 1.9±0.2 | SP | [144] |
| Cu | RT | 500 | 16 | 0.32±0.03 | SP | [106] |
| CuIr | 10 | 5-30 | | 2.1±0.6 | NLD | [145] |
| $Cu_{99.7}Bi_{0.3}$ | ≤10 | 86±17 | 31.25 | -26±11 | NLD | [135] |
| $Cu_{99.5}Bi_{0.5}$ | ≤10 | 45±14 | 19.6 | -24±9 | NLD | [135] |
| $Cu_{99.5}Pb_{0.5}$ | ≤10 | 53±15 | 18.52 | -13±3 | NLD | [135] |
| Ag | | 700 | 15 | 0.7±0.1 | SP | [106] |
| $Ag_{99}Bi$ | ≤10 | 29±6 | 14.7 | -2.3±0.6 | NLD | [135] |
| Mo | 10 | 8.6±1.3 | 2.8 | -（0.8±0.018） | NLD | [146] |
| | RT | / | 4.66 | -（0.05±0.01） | SP | [139] |
| Nb | 10 | 5.9±0.3 | 1.1 | -（0.87±0.2） | NLD | [146] |
| Pd | 10 | 13±2 | 2.2 | 1.2±0.4 | NLD | [146] |
| | RT | 5.5 | 5 | 1.2±0.3 | SP | [147] |
| | RT | 2.0±0.1 | 3.7 | 0.8±0.2 | ST-FMR | [148] |
| Pt | 10 | 11±2 | 8.1 | 2.1±0.5 | NLD | [146] |
| | RT | 1.2 | / | 8.6±0.5 | SP | [149] |
| | RT | 1.4 | / | 12±4 | SP | [141] |
| | RT | 3.4±0.4 | 6 | 5.6±1 | SP | [150] |
| | RT | 2.1±0.2 | 3.6 | 2.2±0.8 | ST-FMR | [151] |
| | RT | 2.1±0.2 | 3.6 | 8.5±0.9 | ST-FMR | [151] |
| | RT | | 2.4 | 4 | SP | [152] |
| | RT | 1.5±0.05 | 0.5-3 | 11±8 | SMR | [153] |
| | RT | 0.9 | 3.2 | 14 | SMR | [154] |
| Ta | 10 | 2.7±0.4 | 0.3 | -0.37±0.11 | NL | [146] |
| | RT | 1.8±0.7 | / | $-2\pm^{0.008}_{0.015}$ | SP | [155] |
| β-Ta | RT | / | / | -（12±4） | ST-FMR | [51] |
| | RT | 1.5±0.5 | / | -（3±1） | SP，SMR | [156] |
| W | RT | 2.1 | 0.55 | -14±1 | SP | [106] |
| | RT | 1.4 | 0.78 | -22 | SMR | [154] |
| β-W | RT | / | 0.38±0.006 | -（33±6） | ST-FMR | [109] |
| Hf（amorphous） | RT | 0.3 | 0.25 | 11（Abs） | SMR | [111] |
| Hf（hexagonal close packed） | RT | 1.3 | 0.64 | 7（Abs） | SMR | [111] |
| Re | RT | 1.5 | 3.57 | 4（Abs） | SMR | [111] |
| PtMn | RT | 0.5±0.1 | 60.1 | 6±1 | SP | [115] |
| IrMn | RT | 0.7±0.2 | 37.1 | 2.5±0.5 | SP | [115] |
| $Ir_{20}Mn_{80}$ | RT | / | / | 2.9±1.5 | ST-FMR | [157] |
| $Ir_{25}Mn_{75}$ | RT | / | 0.7 | 2 | ST-FMR | [158] |
| PdMn | RT | 1.3±0.1 | 44.9 | 1.5±0.5 | SP | [115] |
| FeMn | RT | 1.8±0.5 | 59.8 | 0.8±0.2 | SP | [115] |
| $Bi_2Se_3$ | RT | / | / | 200-350 | ST-FMR | [122] |
| | RT | / | / | 43 | SP | [117] |
| $(Bi_{0.5}Sb_{0.5})_2Te_3$ | 1.9 | / | / | 14000-42500 | SHM | [123] |
| $Bi_{0.9}Sb_{0.1}$ | RT | / | 25 | 5200 | / | [126] |
| $Bi_xSe_{(1-x)}$ | RT | / | 0.0078 | 1862±13 | SMR | [127] |
| $Bi_xSe_{(1-x)}$ | RT | / | 0.0078 | 867±108 | ST-FMR | [127] |
| $LaAlO_3/SrTrO_3$ | RT | / | / | 630 | ST-FMR | [132] |

Table 2

$H_{ext}$ is the assistant magnetic field, $j_c$ is the critical switching current density.

| Materials | $H_{ext}$ (Oe) | $j_c$($10^6$ Acm$^{-2}$) | Reference |
|---|---|---|---|
| Pt(2nm)/Co(0.6nm)/AlOx | 100 | 28 | [52] |
| Pt(5nm)/Co(0.5nm)/Ta(2nm) | 200 | 3.18 | [159] |
| Pt(5nm)/Co(0.5nm)/Ta(8nm) | 200 | 1.51 | [159] |
| Pt(2nm)/MnGa(2.5nm) | 1500 | 50 | [160] |
| Ta(3nm)Pt(5nm)Co(0.6nm)Ta(5nm) | 200 | 2.7 | [161] |
| Ta(3nm)Pt(5nm)Co(0.6nm)/Cr(2nm)/Ta(5nm) | 200 | 2.9 | [161] |
| Ta(2nm)/CoFeB(0.8nm)/MgO(2nm)/$SiO_2$(3) | 400 | 6.55 | [162] |
| Ta(5nm)/MnGa(3nm) | 3000 | 85 | [163] |
| Pd(3nm)/Co(0.6nm)/$AlO_x$ | 2000 | 45 | [164] |
| MgO(4nm)/CoFeB(1.1nm)/Ta(3nm)/CoFeB(0.9nm)/MgO(2nm) | 100 | 0.21 | [165] |
| Hf(5nm)/CoFeB(wedge)/TaOx | 200 | 4 | [166] |
| Hf(5nm)/CoFeB(wedge)/MgO | 0 | 3.5 | [166] |
| W(5nm)/CoFeB(1.2nm)/MgO(1.6nm) | 100 | 4.5 | [167] |
| W(4nm)/Hf(1)/CoFeB(1nm)/MgO(1.6nm)/Ta(1nm) | 3000 | 6.9 | [168] |
| $Bi_2Se_3$(7.4)/CoTb(4.6)/$SiN_x$(3nm) | 1000 | 3 | [124] |
| $Bi_{0.9}Sb_{0.1}$(5nm)/$Mn_{0.45}Ga_{0.55}$(3nm) | 3500 | 1.5 | [126] |
| MgO(2nm)/$Bi_xSe_{(1-x)}$(4nm)/CoFeB(5nm)/MgO(2nm)/Ta | 80 | 0.43 | [127] |

### 3.2 Current-induced antiferromagnets switching by SOTs

Antiferromagnetic materials have zero net magnetization in the ground state and are insensitive to external magnetic fields, so the magnetic order is difficult to manipulate and detect. On the other hand, these properties also make the antiferromagnet a potential candidate for high density and high stability devices. More importantly, the high precession frequency of antiferromagnets offers potential for higher speed of antiferromagnetic devices than that of the ferromagnetic counterparts.

The SOT has proved to be an effective way to modulate the magnetic order in a certain class of antiferromagnets, which was predicted in $Mn_2Au$ by Železný *el al.*, where the two spin sublattices of this antiferromagnet have broken inversion symmetry and form inversion partners [169]. An electrical current generates a local non-equilibrium spin polarization on each Mn sublattice, which averages to zero across the whole sample. The local spin-polarization induces SOTs with staggered components which can be used to control the antiferromagnetic order [170]. Wadley *et al.* [15] experimentally demonstrated the manipulation of antiferromagnetic order in CuMnAs which has a similar structure to $Mn_2Au$. The antiferromagnetic order shown in **figure 6a** was detected using anisotropic magnetoresistance and can be switched reversibly between orthogonal easy axes by the current (see **figure 6b** and **6c**) [15, 171].

Soon after, similar behavior was also observed in $Mn_2Au$ [172-173]. Furthermore, Olejnik *et al*. [174] demonstrated THz electrical writing speed in CuMnAs, compared to the typical GHz writing speed in ferromagnets [175].

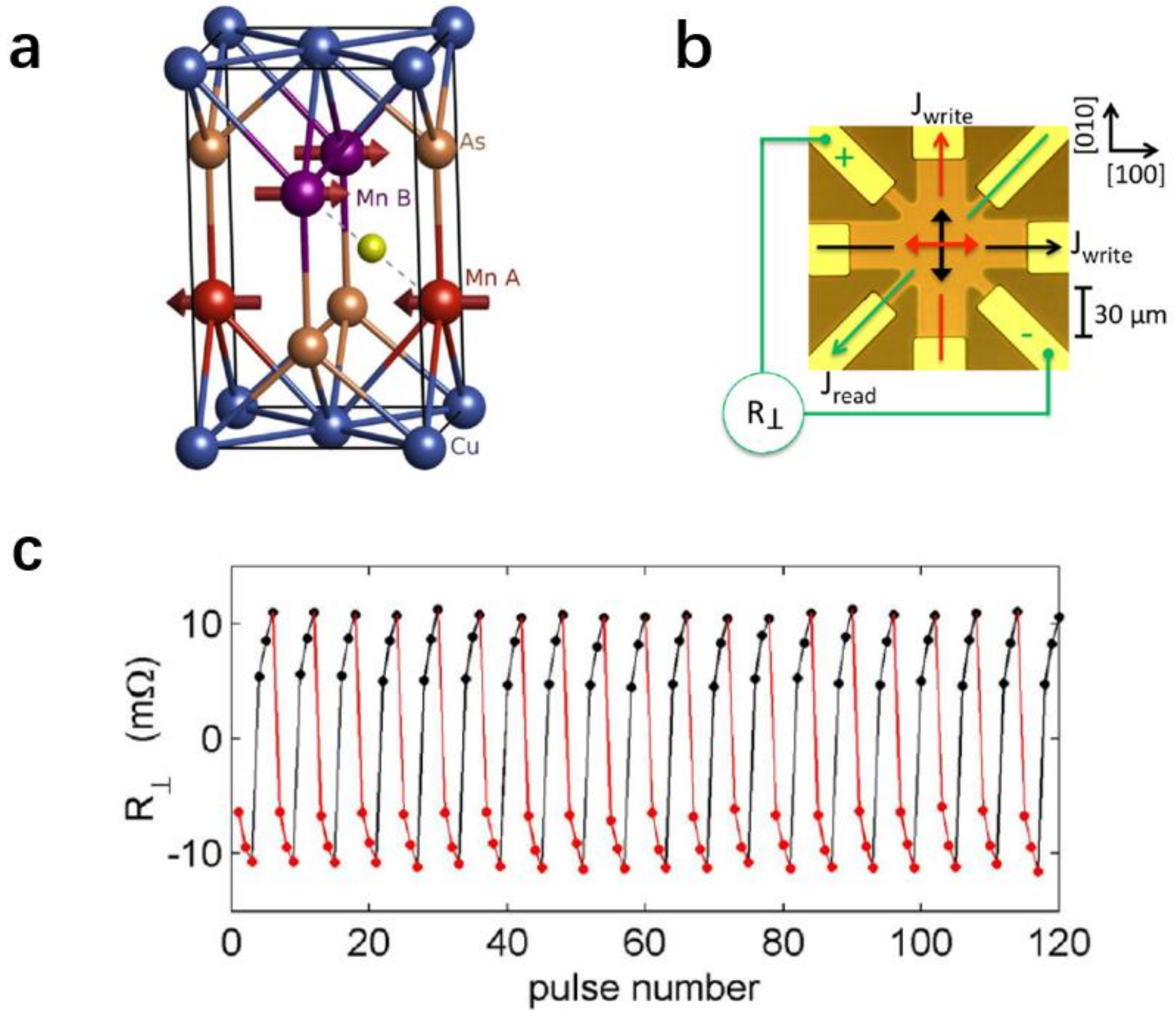

**Figure 6**. Crystal structure of CuMnAs and the magnetization switching of CuMnAs. **a** CuMnAs crystal structure and antiferromagnetic ordering. Mn sublattices A and B are inversion partners around the interstitial position marked by the yellow ball, resulting in opposite sign of local spin polarization; **b** Structure of device; **c** Applying current pulse along [100] crystal axis (black arrow in panel **b** and black point in panel **c**) or along [010] (red arrow in panel **b** and red point in panel **c**) axis and measuring the Hall resistance along [1-10] axis after writing pulse.

Reproduced with permission [15]. Copyright 2016, AAAS.

Manipulation of antiferromagnets by interfacial SOTs in heavy-metal/antiferromagnet heterostructure is still a challenge. SOTs in Ta/IrMn/CoFeB heterostructure were investigated using the second harmonic resistance technique by Reichlová *et al.* [157], revealing a rather complex interplay of antiferromagnetic order and SHEs in the IrMn and Ta layers. Very recently, rotation of the antiferromagnetic order by a damping-like SOT was observed in a Pt/NiO bilayer [176].

Synthetic antiferromagnets, consisting of two ferromagnets with antiparallel coupling, combine the advantages of natural antiferromagnets (high density, high speed and high stability) and ferromagnets (easy to be manipulated and detected). For example, the dipolar coupling from fringing magnetic fields limits storage density in memory devices, but this influence can be eliminated in synthetic antiferromagnets. Furthermore, the domain wall speed in synthetic antiferromagnets can be much larger than that in ferromagnets, as was demonstrated in Co/Ni/Co/Ru/Co/Ni/Co [177]. Furthermore, the synthetic antiferromagnetic exchange coupling enhances the stability of the chiral Neel domain walls, thereby allowing much larger SOTs for the same current density, and therefore higher domain wall velocities. SOT-induced magnetization switching of synthetic antiferromagnets was also reported [178-179].

The manipulation of antiferromagnets by SOTs remains a challenging field. Many obstacles need to be overcome, such as effectively controlling the magnetization in heavy-metal/antiferromagnet structures, detecting the magnetic order with larger signal and with spatial resolution, and accessing the THz switching regime in nanosized device geometries. Even so, this is an area with great potential for future applications.

## 4 Conclusion and outlook

We have reviewed recent progress in manipulation of magnetization by SOTs. The SOTs induced by the bulk and interfacial SOC have been widely used to manipulate the magnetization successfully: efficiently drive domain walls at high speed; deterministically switch the magnetization without external magnetic field through special geometries, exchange bias field and ferroelectric substrate polarization; and tune the magnetic order in antiferromagnets.

To realize applicable memory devices, it is vital to find new, efficient and practical methods to switch the magnetization without external magnetic field. Meanwhile, it is necessary to scale down the device size to improve the storage density. Furthermore, the search for new materials for more efficiently realizing manipulation of magnetization is very important. Materials with large spin Hall angle and small conductivity can decrease the SOTs threshold switching current efficiently.

In modern society, the SOTs effect can be utilized in information processing and storage. The dynamic random access memory (DRAM) based on semiconductor technologies faces the challenge of charge leakage and power consumption increased scaling down of the device sizes. The very low power consumption and high speed offered by SOT-MRAM make it an excellent candidate for the next generation memory, although further optimization is needed.

Spin logic using SOTs has also drawn enormous attention, with the prospect of using spin rather charge to process information. The reconfigurable spin logic and the complementary logic operation have been reported based on the SOTs and VCMA [180-182]. But these attempts are just simply demoed, the further investigations for programmable, ultrafast, multifunctional spin logic devices based on SOTs are extremely necessary.

Imitating the activities of neurons in the brain would markedly increase the processing power for artificial intelligence applications. The possibility of artificial synapses and neural networks based on SOTs has been demonstrated [183-184]. Compared with the complementary metal oxide semiconductor, memristor and memtransistor, the artificial synapses based on SOTs have higher speed, higher durability and higher reliability, which offers a new way to help us better understand this emerging field.

Acknowledgements:

This work was supported by National Key R&D Program of China No. 2017YFB0405700 and 2017YFA0303400. This work was supported also by the NSFC Grant No. 11474272, and 61774144. The Project was sponsored by Chinese Academy of Sciences, grant No. QYZDY-SSW-JSC020, XDB28000000 and K. C. Wong Education Foundation as well.